\documentstyle[art10]{article}
\begin{document}

\begin{titlepage}

\title{Octonions: Invariant Representation of the Leech Lattice }

\author{Geoffrey Dixon\thanks{supported in part by alien invaders.}
\\  Department of Mathematics or Physics \\
 Brandeis University \\
Waltham, MA 02254 \\
 email: dixon@binah.cc.brandeis.edu \\
\and Department of Mathematics \\
University of Massachusetts \\
Boston, MA 02125 \\
 email: dixon@umbsky.cc.umb.edu}

\maketitle

\begin{abstract} The Leech lattice, $\Lambda_{24}$, is represented on
the space of octonionic 3-vectors.  It is built from two octonionic
representations of $E_{8}$, and is reached via $\Lambda_{16}$.  It
is invariant under the octonion index cycling and doubling maps.
\end{abstract}

\end{titlepage}

\section*{1. Introduction.}

My interest in the Leech lattice arose from  {\bf [1]}, in which it
was made clear that  $\Lambda_{24}$ is in some sense the most select
of all lattices.  In {\bf [2]} I made it clear that I felt that the
real division algebras (reals, complexes, quaternions, and
octonions) are the most select of all algebraic objects, and it was
this selectness that inspired my interest in these algebras.  The
same may be said of $\Lambda_{24}$.  However, like many people, I
generally find other people's understandings of mathematical objects
perplexing, and I only gain any degree of understanding myself by
re-representing the objects in a form with which I am comfortable.
That is what I have done here.

The form itself is based on the octonion multiplication I employed
in {\bf [2]}, and in a recent series of papers {\bf
[3-6]}.  The multiplication is one of four for which index
cycling and doubling are automorphisms.  These invariances are
translated to my representation of $\Lambda_{24}$.

I'm going to ask the interested reader to look at {\bf [3-6]} for
background material on the relationship of the octonions to $E_{8}$
and $\Lambda_{16}$.  The notation I employ here, and in those
previous papers, is the same I employed in {\bf [2]}.  The octonion
algebra is denoted {\bf O}.

\section*{2. Three  $E_{8}$'s in $\Lambda_{24}$.}

In {\bf [3]} I employed the following sets to produce
renumberings of the octonion product:

%
%
\begin{equation}
\fbox{$
\begin{array}{clr}
\Xi_{0} = & \{\pm e_{a}\}, & 16 \\ \\
\Xi_{1} = & \{(\pm e_{a}\pm e_{b})/\sqrt{2}: a,b \mbox{
distinct}\}, & 112 \\ \\
\Xi_{2} = & \{(\pm e_{a}\pm e_{b}\pm e_{c}\pm e_{d})/2: a,b,c,d
\mbox{ distinct}, &  \\ \\
&e_{a}(e_{b}(e_{c}e_{d}))=\pm 1\}, & 224 \\ \\
\Xi_{3} = & \{(\sum_{a=0}^{7}\pm e_{a})/\sqrt{8}:
\mbox{ odd number of +'s}  \}, &  \\ \\
& a,b,c,d\in\{0,...,7\}. & 128 \\
\end{array}
$}
\end{equation} \\
The numbers down the right side are the orders of these sets of
octonion units (ie., each of unit norm, so elements of $S^{7}$, the
octonion 7-sphere).  It should be noted that these sets depend on
the octonion product chosen; my choice is that used in {\bf [2-6]}.
\\

{}From these sets we define
%
%
\begin{equation}
\fbox{$
\begin{array}{lll}
{\cal E}_{8}^{even} &=& \Xi_{0}\cup \Xi_{2}, \\ \\
{\cal E}_{8}^{odd} &=& \Xi_{1}\cup \Xi_{3}. \\ \\
\end{array}
$}
\end{equation} \\
Each of these contains 240 elements and is the inner shell (normalized
to unity) of an $E_{8}$ lattice {\bf [4]}.   From ${\cal
E}_{8}^{even}$ we define in ${\bf O}^{3}$:
%
%
\begin{equation}
\fbox{$
\begin{array}{lll}
\Lambda_{24}^{1} &=& \{<A,0,0>,\; <0,A,0>,\; <0,0,A>: \; \; A \in
{\cal E}_{8}^{even}\}.
\end{array}
$}
\end{equation} \\
That is, $\Lambda_{24}^{1}$ consists of all the elements of ${\bf
O}^{3}$ zero in exactly two components, the third component an
element of ${\cal E}_{8}^{even}$.  This set is the first rung in the
ladder to a full Leech lattice, $\Lambda_{24}$ (in particular, we
will end up with an inner shell for $\Lambda_{24}$ consisting of
196560 elements of unit norm in the 24-dimensional space ${\bf
O}^{3}$).  The set
$\Lambda_{24}^{1}$ accounts for $3\times 240 = 720$ elements.

\section*{3. Three $\Lambda_{16}$'s in $\Lambda_{24}$.}

Define
%
%
\begin{equation}
\fbox{$
\begin{array}{lll}
\Lambda_{24}^{2} &=& \{<A,B,0>,\; <0,A,B>,\; <B,0,A>: \; \;
 A,B \in \frac{1}{\sqrt{2}}{\cal E}_{8}^{odd}, \\ \\
&& AB^{\dagger} =\pm\frac{1}{2}e_{a}, \; \; a\in \{0,...,7\}\}.
\end{array}
$}
\end{equation} \\
This constitutes that subset of the inner shell of the full
$\Lambda_{24}$ each element of which has exactly two nonzero
components.  For each pair of components chosen to be nonzero,
there are $16\times 240$ different combinations satisfying the
product condition in (4).  Since there are three ways all together
to fill exactly two components, this subset accounts for $3\times
16\times 240 = 11520$ elements.

In addition, the subset of $\Lambda_{24}^{1}\cup \Lambda_{24}^{2}$
consisting of elements zero in one of the three components (so the
order of such a subset would be $240+240+3840=4320$) is a
representation of the inner shell of the lattice $\Lambda_{16}$ (see
{\bf [5-6]}). \newpage

\section*{4. $\Lambda_{24}$ Inner Shell.}

All that remains is to find the set of elements $\Lambda_{24}^{3}$
of the inner shell of $\Lambda_{24}$ that are nonzero in all three
components.  The order of this set must be
$$
196560 - 720 - 11520 = 184320 = 3\times 16^{2} \times 240.
$$
A rotation of a representation of the Leech lattice developed in {\bf
[1]} gave me a guess as to how to construct a representation of
$\Lambda_{24}^{3}$ consistent with the definitions of
$\Lambda_{24}^{1}$ and $\Lambda_{24}^{2}$.  A desire for it to be as
symmetric as possible led to the representation  below.

Before presenting the representation of $\Lambda_{24}^{3}$, one
last word: in general, if $<A,B,C> \in \Lambda_{24}$ as represented
here, then so are $<\pm A, \pm B, \pm C>$ with all possible sign
combinations, and all six permutations of each of these elements.
In addition, the representation being constructed will be invariant
under both index cycling and index doubling, so given a particular
element many other elements may be easily constructed via these
operations (see {\bf [2]}). \\

All of $\Lambda_{24}^{3}$ can be constructed from the following two
elements in linear combination with $\Lambda_{24}^{1}$ and
$\Lambda_{24}^{2}$:
%
%
\begin{equation}
\fbox{$
\begin{array}{lccc}
\cal{U} \; =& <\frac{1}{4}(1+e_{3}+e_{5}+e_{6}), &
\frac{1}{4}(1+e_{3}+e_{5}+e_{6}), &
\frac{1}{4}(-1+e_{3}+e_{5}+e_{6})+\frac{1}{2}e_{7}>. \\ \\
&\mbox{even}&\mbox{even}&\mbox{odd} \\ \\
& \mbox{with } 1 & \mbox{with } 1 & \mbox{with } 1 \\
\end{array}
$}
\end{equation} \\
%
%
\begin{equation}
\fbox{$
\begin{array}{lccc}  \cal{V} \; =& <\frac{1}{4}(-1+e_{3}+e_{5}+e_{6}),
&
\frac{1}{4}(-1+e_{3}+e_{5}+e_{6}), &
\frac{1}{4}(e_{1}+e_{2}+e_{4}-e_{7})+\frac{1}{2}>. \\ \\
&\mbox{odd}&\mbox{odd}&\mbox{odd} \\ \\ & \mbox{with } 1 &
\mbox{with } 1 & \mbox{w/o } 1 \\
\end{array}
$}
\end{equation}

There are several points to make here.  First, it will be observed
that
$$
\|{\cal{U}}\| = \|{\cal{V}}\| = 1.
$$
Also, since
$$
e_{3}e_{5}=e_{6},
$$
these three octonions form a quaternionic triple.  Hence
%
%
\begin{equation}
\frac{1}{4}(\pm1+e_{3}+e_{5}+e_{6}),\;
\frac{1}{4}(e_{1}+e_{2}+e_{4}-e_{7}) \in \frac{1}{2}\Xi_{2}.
\end{equation}
The first two components of $\cal{U}$ have an even number of
"+" signs and contain the identity ("with 1"; the point of all this
is to develop a pattern).   The third component is the sum of an
element of
$\frac{1}{2}\Xi_{2}$, which has an odd number of "+" signs and
contains the identity, and an element of
$\frac{1}{2}\Xi_{0}$ with a distinct index.  The first two components
of $\cal{V}$ have an odd number of "+" signs and contain the
identity .  The third component is the sum of an element of
$\frac{1}{2}\Xi_{2}$, which has an odd number of "+" signs and
no identity ("w/o 1"), and an element of
$\frac{1}{2}\Xi_{0}$ with a distinct index.

Note also that the elements $\cal{U}$ and $\cal{V}$ are invariant
under index doubling, but  index cycling leads to six other
elements for each.  For example, cycling the indices of $\cal{U}$ by
3 leads to
$$
\begin{array}{lll}   <\frac{1}{4}(1+e_{6}+e_{1}+e_{2}), &
\frac{1}{4}(1+e_{6}+e_{1}+e_{2}),&
\frac{1}{4}(-1+e_{6}+e_{1}+e_{2})+\frac{1}{2}e_{3}>. \\ \\
\end{array}
$$
Index doubling on this leads to
$$
\begin{array}{lll} <\frac{1}{4}(1+e_{5}+e_{2}+e_{4}), &
\frac{1}{4}(1+e_{5}+e_{2}+e_{4}),&
\frac{1}{4}(-1+e_{5}+e_{2}+e_{4})+\frac{1}{2}e_{6}>. \\ \\
\end{array}
$$

Many other unit norm elements may be constructed from $\cal{U}$ and
$\cal{V}$ by adding and subtracting elements of $\Lambda_{24}^{1}$ and
$\Lambda_{24}^{2}$.  Requiring that the result have unit norm,
and that the first two components of the result be the same as those
of $\cal{U}$ leads to the following 16 elements (note that this means
we are modifying $\cal{U}$ with elements of $\Lambda_{24}^{1}$
nonzero in the third component):
%
%
\begin{equation}
\begin{array}{lll}  <\frac{1}{4}(1+e_{3}+e_{5}+e_{6}), &
\frac{1}{4}(1+e_{3}+e_{5}+e_{6}), &
\pm\frac{1}{4}(-1+e_{3}+e_{5}+e_{6})\pm \frac{1}{2}e_{7}>, \\ \\
&& \pm\frac{1}{4}(1-e_{3}+e_{5}+e_{6})\pm \frac{1}{2}e_{2}>, \\ \\
&& \pm\frac{1}{4}(1+e_{3}+e_{5}-e_{6})\pm \frac{1}{2}e_{4}>, \\ \\
&& \pm\frac{1}{4}(1+e_{3}-e_{5}+e_{6})\pm \frac{1}{2}e_{1}>, \\ \\
\end{array}
\end{equation} \\
where each of the $\pm$ pairs are independent.  (Note that this
set of 16 elements is invariant under index doubling.)  In general
we will find that if the first two components are elements of
$\frac{1}{2}\Xi_{2}$, and the overall
element has unit norm, then there will be 16 possible forms for the
third component similar to those shown in (8) above.  Moreover, the
first two such components must differ by multiplication from the
left by some $\pm e_{a},\; a\in\{0,...,7\}$, just as was the case for
the nonzero components of $\Lambda_{24}^{2}$.  Since the order of
$\frac{1}{2}\Xi_{2}$ is 224, and there are three positions for the
peculiar component from $\frac{1}{2}\Xi_{0}+\frac{1}{2}\Xi_{2}$,
this accounts for
$
3\times 16\times 16\times 224
$
new unit elements in $\Lambda_{24}^{3}$. \newpage

Let
$$
\begin{array}{lccc}  \cal{V}' \; =&
<\frac{1}{4}(-1+e_{3}+e_{5}+e_{6}), &
\frac{1}{4}(-1+e_{3}+e_{5}+e_{6}), &
\frac{1}{4}(-e_{1}-e_{2}-e_{4}+e_{7})-\frac{1}{2}>, \\ \\
\end{array}
$$
which is the same as $\cal{V}$ with the sign of the third component
changed.  Take the difference,
%
%
\begin{equation}
\fbox{$
\begin{array}{lccc}
\cal{W} \; =\; \cal{U}-\cal{V}' \; =& <\frac{1}{2}, &
\frac{1}{2}, &
\frac{1}{4}(1+e_{1}+e_{2}+e_{3}+e_{4}+e_{5}+e_{6}+e_{7})>. \\ \\
\end{array}
$}
\end{equation} \\
As was true of $\cal{U}$ and $\cal{V}$, the element $\cal{W}$ is
index doubling invariant.  And it has unit norm.  Maintaining the
unit norm property, and modifying only the third component with
elements of $\Lambda_{24}^{1}$, leads to the following 16 variations:
%
%
\begin{equation}
\fbox{$
\begin{array}{lccc}
\cal{W} \; =\; \cal{U}-\cal{V}' \; =& <\frac{1}{2}, &
\frac{1}{2}, &
\pm\frac{1}{4}(1+e_{1}+e_{2}+e_{3}+e_{4}+e_{5}+e_{6}+e_{7})>, \\ \\
&&&\pm\frac{1}{4}(1-e_{1}-e_{2}-e_{3}+e_{4}-e_{5}+e_{6}+e_{7})>, \\ \\
&&&\pm\frac{1}{4}(1+e_{1}-e_{2}-e_{3}-e_{4}+e_{5}-e_{6}+e_{7})>,\\ \\
&&&\pm\frac{1}{4}(1+e_{1}+e_{2}-e_{3}-e_{4}-e_{5}+e_{6}-e_{7})>,\\ \\
&&&\pm\frac{1}{4}(1-e_{1}+e_{2}+e_{3}-e_{4}-e_{5}-e_{6}+e_{7})>,\\ \\
&&&\pm\frac{1}{4}(1+e_{1}-e_{2}+e_{3}+e_{4}-e_{5}-e_{6}-e_{7})>,\\ \\
&&&\pm\frac{1}{4}(1-e_{1}+e_{2}-e_{3}+e_{4}+e_{5}-e_{6}-e_{7})>,\\ \\
&&&\pm\frac{1}{4}(1-e_{1}-e_{2}+e_{3}-e_{4}+e_{5}+e_{6}-e_{7})>\\ \\
\end{array}
$}
\end{equation} \\
(this is another set closed under index doubling, and in this case,
also index cycling).  In general, for any element whose first two
components are drawn from $\frac{1}{2}\Xi_{0}$ (and all such
combinations are occur), there will be 16 possible third components
(again, assuming overall unit norm), each an element of
$\frac{1}{2}\Xi_{3}^{\dagger}$.  So that means 16 first components
from $\frac{1}{2}\Xi_{0}$, and for each of these 16 second
components from $\frac{1}{2}\Xi_{0}$, and for each combination 16
third components from $\frac{1}{2}\Xi_{3}^{\dagger}$.  Taking
into account permutations of the $\frac{1}{2}\Xi_{3}^{\dagger}$
element to the first and second components, we have accounted for
$3\times 16\times 16\times 16$
new unit elements of $\Lambda_{24}^{3}$.  And that's all there are.
So the order of $\Lambda_{24}^{3}$ is, as expected
$$
3\times 16\times 16\times (16+224) = 3\times 16\times 16\times 240.
$$
The combined elements of
$$
\Lambda_{24}^{1}\cup\Lambda_{24}^{2}\cup\Lambda_{24}^{3}
$$
constitute an inner shell of a representation of the Leech lattice,
$\Lambda_{24}$. \\

The interested reader should now have enough information to be able
to get a feeling for how this representation works.  I am finding it
considerably easier to play with than the other representations
I've found.  Further developments will be revealed as they become
available (as, presumably, will be my reason for pursuing this line
of questioning).
\\

I'd like to acknowledge several electronic conversations with Tony
Smith, who maintains a fascinating Web site at Georgia Tech:
www.gatech.edu/tsmith/home.html

\end{document}